\documentclass[twocolumn,showpacs,preprintnumbers,amsmath,amssymb,floatfix]{revtex4}
\usepackage[utf8]{inputenc}
\usepackage{graphicx}% Include figure files
\usepackage{dcolumn}% Align table columns on decimal point
\usepackage[usenames]{color}

\begin{document}

\title{A perfect fluid model for compact stars}

\author{Gabino Estevez-Delgado}
\email{gestevez.ge@gmail.com} \affiliation{Facultad de Qu\'imico Farmacobiolog\'ia de la Universidad
Michoacana de San Nicol\'as de Hidalgo, Tzintzuntzan No. 173, Col.
Matamoros, C.P. 58240, Morelia Michoac\'an, M\'exico.}
\author{Joaquin Estevez-Delgado}
\email{joaquin@fismat.umich.mx} \affiliation{Facultad de Ciencias
F\'isico Matem\'aticas de la Universidad Michoacana de San Nicol\'as de
Hidalgo,  Edificio B, Ciudad Universitaria,  CP 58030,  Morelia Michoac\'an, M\'exico.}
\author{Nadiezhda Montelongo García}
\email{nmontelongo@itspa.edu.mx}
\affiliation{ Instituto Tecnológico Superior de Pátzcuaro,Av. Tecnológico No.1 Zurumutaro, P\'atzcuaro Michoac\'an, M\'exico\\
}
\author{ Modesto Pineda Duran}
\email{mpinedad@itstacambaro.edu.mx} \affiliation{Instituto Tecnológico Superior de Tacámbaro, Av. Tecnológico No 201, Zona el Gigante, C: P. 61650,  Tacambaro Michoac\'an, M\'exico.}
\date{\today}
\begin{abstract}
\noindent
In the framework of Einstein's the theory of general relativity we present a new  interior solution with a perfect fluid, this is constructed from the proposal of a gravitational redshift factor. The geometry is regular and its density and pressure  are monotonic decrescent functions, furthermore the sound speed
is smaller than the light speed and monotonic crescent. The solution depends on a parameter $w\in(0,2.0375509325]$ related to the compactness of the star $u=GM/c^2R$, the maximum value $u=0.2660858316$ which allow to describe compact stars like quark stars or neutron stars.
Although there is a diversity of stars for which the model can be used, we only apply this solution to describe the interior of a neutron star PSR J0348+0432.
According to the observations, it is known that its mass $ M=(2.01\pm 0.04 )M_\odot $ and its radius is between $ 12,062Km $ and $ 12,957 Km $, so the value of the compactness is in the range $ u\in [0.2244845,0.2509338] $.
In addition to the decreasing behavior of the mentioned pressure and density functions,
 the results are consistent with the density values range typical of neutron stars and the maximal central density of the star result to be $1.283818 \times 10^{18} Kg/m^3$. \\

\noindent
\end{abstract}

\maketitle
\section{Introduction}
The theoretical development on the constituents of the interior of the stars have allowed a better understanding of these, as a consequence of  the advances in particle physics \cite{GeorgParticulas1,WeberParticulas1}. For decades it has worked about white dwarf stars, neutron stars and quark stars \cite{ChandrasekharBook,Fechner,Glendenning,NaNeutron,WeberQuarks,WeberQuarks1}. The advances in the theoretical direction are limited in part by the observational difficulty that allows to decide on the model that adequately describes the interior of compact stars.
In addition, some observations focus on the determination of the mass and radius of the stars, although there are also novel proposals such as the  based on  gravitational wave observations that allow to determine the equation of state of matter of quarks \cite{Hajime} and this allows to give a possible equation of state for compact stars.
\\
On the other hand, the approach that some researches have proposed to understand the interior of compact stars is based on the construction of exact solutions to the Einstein's equations with matter given by an anisotropic perfect fluid or charged\cite{Newton,Nematollah,Sharif} . This approach has allowed to obtain some general conclusions for stellar models.This approach showed that the ratio of compactness that limits the value of the possible mass and radius is given by  $u=\frac{GM}{c^2R}<\frac{4}{9}$ \cite{u},  where $R$  denotes the radius and $M$ the mass. This relation is valid if the matter inside the star is described by a perfect fluid with decreasing monotonous pressure and density.
\\
Some models proposed as a result of the solution of the Einstein's equations with the source of matter of a perfect fluid and  static and spherically
  symmetric spacetime have been used to describe the behavior of stars \cite{estrellas1,estrellas2,estrellas3,estrellas4}, although most of the solutions presented do not satisfies conditions that make them physically acceptable \cite{Delgaty}.
An analysis of the behavior of the known solutions until before 1998, it has shown that of 127 solutions analyzed only 16 of them pass the conditions that make them physically acceptable and of these only 9 comply that their speed of sound is a decreasing monotone function as a function of the radial distance \cite{Delgaty}. This shows the difficulty of building analytical solutions that are physically acceptable, i.e.,  solutions with regular geometry absent from the event horizon and with regular functions  of density and pressure monotonous decreasing.
There are works where some models have been constructed with physically acceptable perfect fluid that describe compact stellar objects
\cite{fp1,fp2,fp3}. In addition to the exact solutions for stellar models, numerical solutions have been proposed in which a specific form of the state equation that describes the interior of the stars is supposed, the restriction of the state equation makes it more difficult to build analytical solutions by what is chosen by the use of numerical methods to describe the behavior of the interior of the stars \cite{numericos1,numericos2,numericos3}.
\\
There are fewer solutions to Einstein's equations with a perfect fluid than solutions with anisotropic  or charged fluid. One of the reasons is that solutions for the last two cases can be generated from a solution with perfect fluid and the choice of the anisotropy shape function or the charged function, although they must satisfy specific properties, are not unique, while another class of anisotropic or charged solutions are not the result of the generalization of solutions with perfect fluid \cite{Krori}.
\\
In addition, solutions of Einstein' equations associated with a perfect fluid in a static and spherically symmetric spacetime  does not imply that it can have a physically acceptable model. However, from these solutions in some cases a physically acceptable solution it was built for the case of a fluid with anisotropic pressures \cite{GestevezEJPC}, in this work it has been shown the importance of anisotropy in stellar models and have been proposed more models than in the case of perfect fluid.  The relevance of the anisotropy has led to the construction of anisotropic solutions, some of them start from a seed of solutions with perfect fluid \cite{Guptaf}, solutions with a geometric restriction \cite{Newton1, Newton2} and the cases in which  equation of state is giving by   $ P = P (\rho) $ and in some cases a physically acceptable solution with perfect fluid is not recovered when the anisotropy factor is zero \cite{Bhar1, Maurya, Bhar2}.
\\
For several years the approach of charged solutions has also attracted attention and in recent times the number of works on this topic them have intensified.
There are works of regular solutions that generalize to spacestime as the internal solution of Schwarzschild \cite{GuptaKumar}  just like the singular solutions \cite{Saibal}, although most of the recent work is more focused on stellar models that are regular
\cite{Mamta,Peadhan,Basanti}. For the more general case in which there is an anisotropic fluid with charge, solutions have also been proposed as well as their applications to different stars for which there are observational data on their mass and radius \cite{Newtonmix,Neeraj,Mohammad,Esculpi}.
\\
In this work, following the idea of some research reports in which exact interior solutions have been built for a static and spherically symmetrical spacetime with perfect fluid, starting from a new form of the gravitational redshift factor, it is different from the previously proposed \cite{Delgaty} given as a function of the form $(1+ar^2 )^n$, we present a new stellar model within the framework of general relativity suitable to represent compact stars.
\\
Although the presented solution has its relevance by itself, because it is applicable to describe observed stars and the values of the hydrostatic variables are consistent with the expected orders of magnitude, it is worth mentioning that having a stellar solution with perfect fluid takes us to a series of theoretical implications and increases the possibilities of a better understanding of stellar interior behavior through the analysis of the same solution as well as of variants or generalizations in which anisotropies or models are contemplated loaded either described by a perfect fluid or anisotropic.
\\
We could also obtain new models with perfect fluid, taking this as seed, and through the application of proposed theorems generate new solutions with perfect fluid \cite{Lake1, Petarpa1, Petarpa2, Petarpa3}.
\\
Giving rise to a diversity of research works that could be developed in the future. In the next section we give the equations that describe the interior of a star and from the assignment of the explicit form for the function of gravitational redshift we build the solution.
\\
The section \ref{III} focuses on the algebraic and graphical analysis of the solution and restriction of the parameters to determine a physically acceptable solution.
\\
In the section \ ref {IV} for the data of the neutron star PSR J0348 + 0432 \ cite {Antoniadis, Zhao}. We finalize this paper with the conclusions section where the future works that arise as a result of the proposal presented in this paper are also discussed.
\section{The model  \label{II} }
For the construction of our model, we suppose that it is described by a static and spherically symmetric spacetime, so the metric is expressed in the form \cite{Wald}:
\begin{eqnarray}
ds^2\!=\!-y^2(r)dt^2+\frac{dr^2}{B(r)}
   +\!r^2 (d\theta ^2+\sin ^2{\theta}\,d\phi ^2),
\label{elementodelinea}
\end{eqnarray}
where the metric functions $y(r)$ y $B(r)$ describe the geometry of the interior of the star with matter given by a perfect fluid, i.e. the energy-moment tensor is given:
\begin{equation}
 T_{\mu\nu}= (P+c^2\rho) u_{\mu}u_{\nu}+P g_{\mu\nu}, \label{fluido}
\end{equation}
where $\rho$ represents the energy density and  $P$ the pressure and $c$ the speed of light. From the  Einstein's  field equations
$G_{\mu\nu}=kT_{\mu\nu},$ where $k =\frac{8\pi G}{c^4} $ is the coupling constant, then \cite{Wald}:
\begin{eqnarray}
kc^2\rho\! &\!=\!& \!-\frac {B'}{r}+\frac{1-B}{r^2},  \label{rho} \\
k{\it P} \!&\!=\!&\! \frac{2By'}{ry }
-\frac{1-B}{{r}^{2}} , \label{Pr} \\
k{\it P} \!&\!=\!&\!{\frac { ( ry''+y')B }{ry} }- {\frac{( ry'+y)B'}{2ry }},\qquad \label{Pt}
 \end{eqnarray}
where $'$ denotes the derivative with respect to the coordinate $r$.
From the combination of the equations (\ref{rho})-(\ref{Pt}), then
\begin{equation}
{\it P}'=-{\frac { \left( {\it P} +c^2\rho  \right) y'}{y }},\label{conservacion}
%T_{\mu\nu}= (P+c^2\rho) u_{\mu}u_{\nu}+P g_{\mu\nu}, \label{fluido}
\end{equation}
\noindent So the effective system of equations are the three equations  (\ref{rho})-(\ref{Pt})
with the four functions to be determined, so we can impose an additional equation or constraint, which may well be an equation of state $P=P(\rho)$ or some other relation as the assignment of a metric function, which is the way we will proceed.
To solve the system, note that by subtracting the equations (\ref{Pr}) and (\ref{Pt}) we get a differential equation that only involves the metric components and their derivatives:
\begin{equation}
( ry'+y)rB'+2( r^2y''-ry'-y)B+2y=0,
\label{Em}
\end{equation}
the integration of this can be done
assuming the shape of one of the metric functions $y$ or $B$ and solving for the other.
The convenience of assigning a specific form of the gravitational redshift factor $y$ as a starting point for the construction of solutions to the stellar model system with perfect fluid has been used previously \cite{Gestevez}, since the equation (\ref{Em}), once assigned a form of $y$, it turns out to be a non-homogeneous ordinary differential equation of the first order and this can facilitate the construction of new exact solutions.
The relation between the gravitational redshift factor $y$ and the function of gravitational redshift
$z(r)$ is given by $z(r)=y(r)^{-1}-1$, that by the conditions of continuity of the inner solution and the external solution described by the Scwarzschild metric on the surface of the star the gravitational redshift value is $z=1/\sqrt{1-2GM/c^2R}-1$.
In our case we take the gravitational redshift factor given by the following expression
$$
y \left( r \right) =\frac{C(  5+{4}a{r}^{2
} )}{ \sqrt{1+a{r}^{2} } },
$$
where $C$ and $a$ are  constant, substituting this in (\ref{Em})
\begin{eqnarray}
(\!\!&\!\!\!5\!\!\!&
+12a{r}^{2}+8{a}^{2}{r}^{4})(1+a{r}^{2})rB'
-( 10+28\,a{r}^{2} \qquad\qquad
\nonumber \\
&\!\!+\!\!&
28{a}^{2}{r}^{4}+
16{a}^{3}{r}^{6}) B +2( 5+4a{r}^{2} ) ( 1+a{r}^{2}) ^{2}=0,
\label{Em1}
\end{eqnarray}
then
\begin{equation}
B (r) \!=\frac{( 5+11a{r}^{2}+6\,{a}^{2}{r}^{4}-4
a{r}^{2} S(r) ) ( 1+a{r}^{2}) }{5+12a{r}^{2}+8{a}^{2}{r}^{4}},
\label{B}
\end{equation}
where
$$
S\left( r \right) =
\frac{( 1+a{r}^{2}) ^{2}
 \left[A+\arctan\!{\mbox h}\left({\frac {1+2a{r}^{2}}{\sqrt {5+12a{r}^{2}+8{a}^{2}{r}^{4}}}} \right)  \right]}
 {\sqrt {5+12\,a{r}^{2}+8\,{a}^{2}{r}^{4}}}.
$$
 with $A$ the integration constant. When we get $(y,B)$ of (\ref{rho}) and (\ref{Pr}) we obtain density and pressure respectively. In the next section, we discuss the
conditions that must be met for a solution of Einstein's equations with perfect fluid to be physically acceptable.
\section{Physicals conditions}
\label{III}
\noindent For an inner solution of Einstein's equations with perfect fluid associated with compact objects, the model must satisfy the following conditions\cite{Delgaty}:
\begin{itemize}
\item
The solution must not have singularities, i.e.,  for  $0\leq r\leq R$ the curvature scalars must be regular and the metric functions
 $(y^2,B)$, the density and central pressure are positive.
\item
The pressure and density must be positive and monotonous decreasing functions as a function of radial distance, with its maximum value in the center, so in particular in the
origin:
$$
P(0)>0,\qquad
\left.\frac{dP}{dr}\right|_{r=0}=0, \qquad
\left.\frac{d^2 P}{dr^2}\right|_{r=0}<0,
$$
$$
\rho(0)>0,\qquad
\left.\frac{d\rho}{dr}\right|_{r=0}=0, \qquad
\left.\frac{d^2 \rho}{dr^2}\right|_{r=0}<0,
$$
while for $r\neq 0$ $\rho'<0$ y $P'<0$.
\item
The condition of causality must not be violated, i.e. the magnitude of the speed of sound must be less than the speed of light
$$
0\leq v^2=\frac{\partial }{\partial \rho}P(\rho)
= \left.\frac{dP}{dr} \right/\frac{d \rho }{dr}
\leq c^2,
$$
and additionally we will impose that the speed of sound is a monotonous function decreasing towards the surface.
\item
For the stability of the solution, in the relativist case, it is required that
the adiabatic index $\gamma=\frac{c^2\rho+P}{ c^2P}\frac{dP}{d\rho}=\frac{c^2\rho+P}{P}\frac{v^2}{c^2}>\frac{4}{3}$,
\cite{Heintzmann,Shapiro}.
The adiabatic index in the same way as the magnitude of the speed of sound can be obtained through the chain rule if known $P=P(r)$ y $\rho=\rho(r)$.
\item
There must be a region $r=R$, the surface of the star, where the pressure is  $P(R)=0$.
\item
About the border $r=R$ the internal metric and the external metric must be continuous. The exterior geometry is described by the Schwarzschild metric:
\begin{eqnarray}
ds^2\! &\!=\!&-\left(1-\frac{2GM}{c^2r}\right)\,dt^2+\left(1-\frac{2GM}{c^2r}\right)^{-1}dr^2
\nonumber \\
& &\, +\,r^2 (d\theta ^2+\sin ^2{\theta}\,d\phi ^2),
\quad r\geq R,
\label{elementodelineae}
 \end{eqnarray}
where $M$ is the mass of the star, $G$ is the gravitational constant and $c$ the speed of light.
\end{itemize}
These basic requirements allow to determine which interior solution can be useful as a model for the description of some compact object.
\section{Analysis of the solution } \label{IV}
\begin{figure}[h!]
 	\centering
 	\includegraphics[scale=0.57]{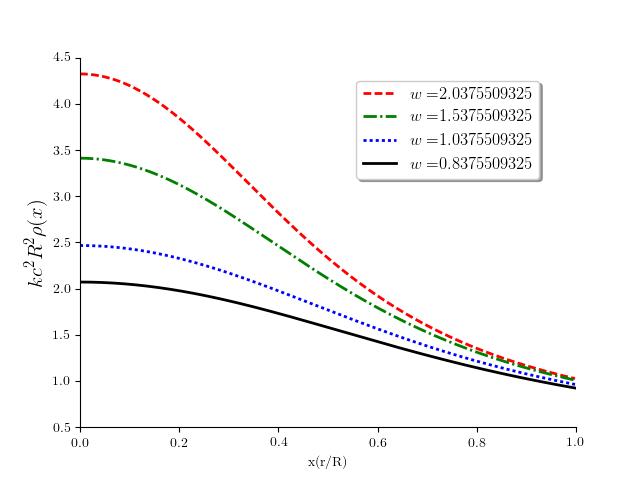}
 	\caption{Density behavior}
 	\label{densidad}
 \end{figure}

\begin{figure}[h!]
 \centering
     	\includegraphics[scale=0.57]{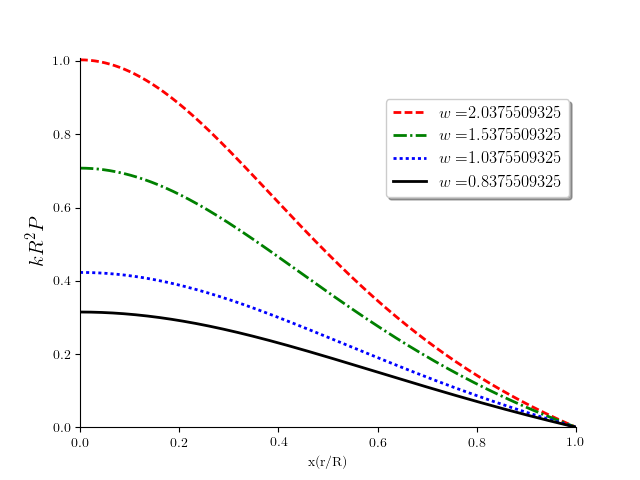}
 	\caption{Pressure for different values of the parameter $w$ }
 	\label{presion}
 \end{figure}

\begin{figure}[h!]
	\centering
	  \includegraphics[scale=0.57]{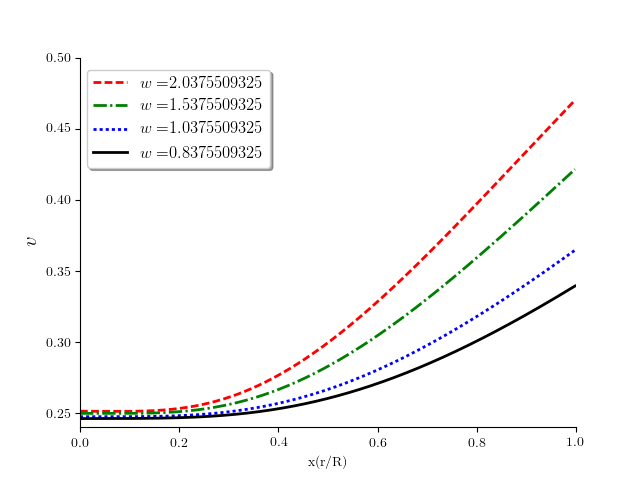}
 	\caption{Speed of sound}
 	\label{velocidad}
 \end{figure}

\begin{figure}[h!]
 \centering
  \includegraphics[scale=0.57]{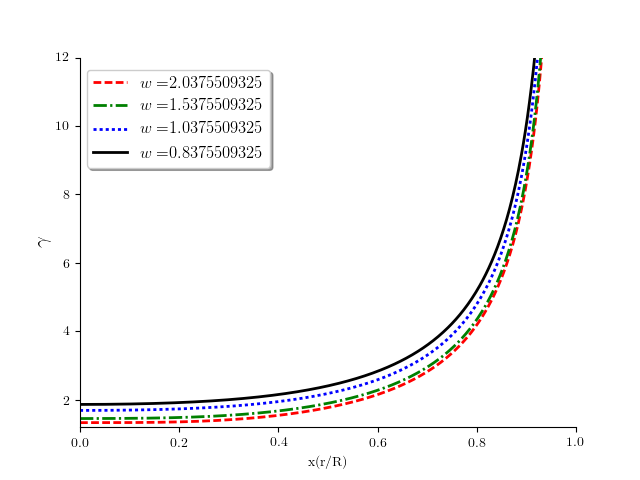}
 \caption{Graph of the behavior of the adiabatic index}
 	\label{pfindice}
 \end{figure}

Once that the metric functions are known, pressure and density are obtained by direct substitution of $(y, B)$  in (\ref{rho})  and (\ref{Pr}), we get
\begin{eqnarray}
\!\!\!\!\rho(r)=\!\! &\!-\!& \!\!
\frac{a \left( 60+265a{r}^{2}+486{a}^{2}{r}^{4}+424{a}^{3}{r}^{6}+144
{a}^{4}{r}^{8} \right) }{ k{c}^{2}\left( 5+12a{r}^{2}+8{a}^{2}{r}^{4}
 \right) ^{2}}
\nonumber \\
&\!\! & \!
+\,\frac{12a \left( 5+15 a{r}^{2}+16{a}^{2}{r}^{4}+8{a}^{3}{r}^{6} \right) S\left( r \right) }
{ k{c}^{2} \left( 5+12 a{r}^{2}+8 {a}^{2}{r}^{4} \right) ^{2}},
\label{rhof} \\
{\it P(r)} \!\!&\!=\!&\! \!\frac {a
 \left( 50+167\,a{r}^{2}+190\,{a}^{2}{r}^{4}+72\,{a}^{3}{r}^{6}
 \right) }{ k\left( 5+12\,a{r}^{2}+8\,{a}^{2}{r}^{4} \right)  \left( 5+
4\,a{r}^{2} \right) }
\nonumber \\
&\!\! & \!
-\frac {4a \left( 5+15\,a{r}^{2}+12\,{a}^{2}
{r}^{4} \right) S \left( r \right) }{ k\left( 5+12\,a{r}^{2}+8\,
{a}^{2}{r}^{4} \right)  \left( 5+4\,a{r}^{2} \right) }.
\label{Prf}
\end{eqnarray}
Determined the pressure and density, the calculation of the speed of sound is given:
%$$
%v^2=\frac{\partial }{\partial \rho}P(\rho)
%= \left.\frac{dP}{dr} \right/c^2\frac{d \rho }{dr}
%$$
\begin{equation}
\frac{{\it v^2}(r)}{c^2}\!=\frac{\left( 5+12a{r}^{2}+8{a}^{2}{r
}^{4} \right)  \left( N_1S\left( r \right)
-N_2 \right) ( 3+4a{r}^{2}) }
{\left( 5+4a{r}^{2} \right) ^{2} \left(N_3S\left(r\right)
-N_4(1+ar^2)\right) },
\end{equation}
where
\begin{eqnarray}
{\it N_1}(r)&\!\!=\!& \!\!
4(25+75\,a{r}^{2}+70\,{a}^{2}{r}^{4}+24\,{a}^{3}{r}^{6}),
\nonumber \\
{\it N_2} (r)&\!\!=\!& \!\!
(1+ar^2)(25+40a{r}^{2}+28{a}^{2}{r}^{4}+16{a}^{3}{r}^{6}),
\nonumber \\
{\it N_3} (r)&\!\!=\!& \!\!12(
25+115a{r}^{2}+166{a}^{2}{r}^{4}+88{a}^{3}{r}^{6}+16{a}^{4}{r}^{8}),
\nonumber \\
{\it N_4} (r)&\!\!=\!& \!\!
175+480a{r}^{2}+372{a}^{2}{r}^{4}+96{a}^{3}{r}^{6}+32{a}^{4}{r}^{8}.
\nonumber
\end{eqnarray}
The imposition of the conditions mentioned in the previous section that guarantee that the solution is physically acceptable sets the range of constants $(a,A)$. Regarding the conditions to impose so that the solution is physically acceptable, we have that the metric functions are regular and positive and the curvature scalars are regular. For the rest of the conditions we will start by evaluating the density and the pressure at the origin and indicating the inequality that must satisfy:
\begin{equation}
k{c}^{2}\rho \left( 0 \right) ={\frac {12}{5\sqrt {5}}}\!\left[ -
\sqrt {5}+A+{\rm arctanh}\, {5}^{-\frac{1}{2}} \right] a>0,
\label{rho0}
\end{equation}
\begin{equation}
kP( 0 ) ={\frac {2}{5\sqrt {5}}}\! \left[ 5\,\sqrt {5}-2(A+{\rm arctanh}\,{5}^{-\frac{1}{2}})  \right] a>0,
\label{Pr0}
\end{equation}
then
\begin{equation}
k \left( \rho \left( 0 \right) {c}^{2}+3\,\Pr \left( 0 \right)  \right) ={\frac {18\,a}{5}}>0,
\label{aa}
\end{equation}
of this expression we get that $a>0$. The first derivative of the pressure and density evaluated at the origin is zero, while the second derivatives of the pressure and density at the origin are
\begin{equation}
kP''(0)={\frac {6}{25\sqrt {5}}}\! \left[\sqrt {5}-4(A+{\rm arctanh} \,{5}^{-\frac{1}{2}}) \right] \!{a}^{2}\!<0,
\label{Pr20}
\end{equation}
\begin{equation}
k{c}^{2}\rho''(0)={\frac{2}{5\sqrt {5}}}\left[7\,\sqrt {5}-12(A+{\rm arctanh} \,{5}^{-\frac{1}{2}}) \right] {a}^{2}<0.
\label{rho20}
\end{equation}
While determining the speed of sound at the origin
$$
{\frac {  {v^2}(0) }{{c}^{2}}}={\frac { 3[\sqrt {5}-4(A+{\rm arctanh} \,{5}^{-\frac{1}{2}})]}{5[7\,\sqrt {5}-12(A+{\rm arctanh} \,{5}^{-\frac{1}{2}})] }}<1.
$$
From the condition that the speed of sound must be less than that of light and that the second derivative of the density evaluated at the origin must be negative we have
\begin{equation}
\frac{2\sqrt {5}}{3} -{\rm arctanh} \frac{1}{\sqrt {5}}  <A,
\label{v0}
\end{equation}
\begin{equation}
\sqrt {5}-{\rm arctanh} \frac{1}{\sqrt {5}}<A<\frac{5\sqrt {5} }{2}-{\rm arctanh} \frac{1}{\sqrt {5}}.
\label{Ad}
\end{equation}
This interval was obtained only using the conditions in the center, its relation with the constant $a$ is obtained by imposing that the pressure is annulled on the surface of the star, located in $r=R$, then:
\begin{eqnarray}
A&\!=\!& \!
{\frac { \left( 50+167w+190{w}^{2}+72{w}^{3}
 \right) \sqrt {5+12w+8w^{2}}}{ 4\left( 5+15w+12w^{2}
 \right)  \left( 1+w \right) ^{2}} }
 \nonumber \\
&\!\!& \!\!-{\rm arctanh} \left( {\frac {1+2w}{\sqrt {5+12w+8w^{2}}}},
 \right)\label{A}
\end{eqnarray}
where $w=aR^2>0$. A graphical analysis of the density and pressure behavior in the interior shows that these are definite positive and monotonous decreasing functions. The speed of sound
  is positive less than the speed of light and monotonous
increasing as a function of the distance to the center of the star, a stronger  condition on the behavior of the speed of sound requires that this be a
decreasing monotone function, in our case the solution does not satisfy this condition \cite{Delgaty}.
Another characteristic that should be imposed on the model, given that this is relativistic, is that the adiabatic index satisfies\cite{Heintzmann}:
$$
\gamma=\frac{\rho+P}{P}\frac{\partial P}{\partial \rho}>\frac{4}{3}
$$
this is the condition that most restricts the parameter range $w$, then $w\in(0,2.0375509325]$.
To know the type of compact objects that can be described with this model we calculate the compactness ratio $u=GM/c^2 R$, where $M$ is the mass of the object and $R$ the radio.
From the continuity of the metric on the surface, the inner metric (\ref{elementodelinea}) and the external metric (\ref{elementodelineae}), then
$$
y^2(R)=1-\frac{2GM}{c^2R}\qquad,
B(R)=1-\frac{2GM}{c^2R}=1-2u.
$$
From the first of these equalities we obtain the value of the constant $C$, while of the second equality implies that the value of compactness is
$$
u=\frac{1}{2}(1-B) ={\frac {w \left(3+ 4\,w\right) }{5+15\,w+12\,{w}^{2}}}.
$$
this function is monotonously increasing so, given the interval
$w\in(0,2.0375509325]$, its maximum value occurs for $w= 2.0375509325$ and this is $u=0.2660858316$ that allows to describe compact stars.
Now we make use of graphic representations to show the behavior inside the star and this is represented in the figures
\ref{densidad}-\ref{pfindice}.
Although we have chosen some specific values of the parameter $w$, its behavior is similar for other values of $w$ in the interval for which the solution is physically
acceptable.
\noindent
To graphically represent the behavior of density, we define the dimensionless variable $\overline{\rho}=kc^2R^2\rho$ as a function of $x=r/R$ for different values
of the parameter $w$. On the graph \ref{densidad} its monotonous and decreasing behavior with respect to the radial or dimensionless coordinate is shown $x$.

 \noindent The dimensionless function for pressure is now given by $\overline{P}=kR^2P$ and its decreasing monotonic behavior as a function of radial distance is described in the figure \ref{presion}. Note that pressure and density increase their values if the value of the parameter is greater.

 From the figure \ref{velocidad}, it is observed that the speed of sound is a growing monotonous function, taking its maximum value at the border, i.e.  $v^2=0.4707785807 c^2$
for the case in which the parameter $w=2.0375509325$. The graph shows that for smaller values the speed of sound is less than this value.

The condition that reduced the parameter range $w$ is the adiabatic index its lowest value$\gamma=4/3$ occurs in the center to $w=2.0375509325$ and for higher values of
parameter $w$ the adiabatic index is higher, this is shown in the graph \ref{pfindice}.

\section{Application of the model to the Neutron Star PSR J0348+0432}\label{V}
The compactness value determined for the model allows to describe compact stars such as neutron stars. As an application and consistency of our model we will apply it
to the neutron star PSR J0348+0432 that according to the observational data and its analysis it is known that its radius is between $12.062Km$  and $12.957 Km$ and its mass
$M=2.01\pm 0.04 M_\odot$ \cite{Antoniadis,Zhao}. Can be seen from the figure \ref{densidad}  the form of the density function is independent of the value of the parameter $w$, while of the figures \ref{presion}- \ref{pfindice} we note that also in the case of the functions of density, pressure, speed of sound and adiabaticolala form of the respective functions does not depend on the parameter $w$ so the range of possible values in each case is between the values in the center and the border.
\\
For the neutron star PSR J0348+0432 with the minimum compactness $u=0.2244845$, that occurs when the radius is maximum $R=12.957 Km$ and the minimum mass $M=1.97 M_\odot$, in the first row of the table \ref{Tabla} we present the intervals for density, pressure, speed of sound and adiabatic index. In the second row of the table\ref{Tabla} we give the respective intervals for maximum compactness $u=0.2509338$ that occurs when the radius is minimal
$R=12.062Km$ and the mass is maximum $M=2.05 M_\odot$. It is convenient to note that the adiabatic index tends to infinity on the surface of the star, while the pressure on the surface of the star vanishes.
\begin{table}[htbp]
\begin{center}
\begin{tabular}{|c|c|c|c|c|c|}
\hline
$\!\!\rho_c\!$&$\!\!\rho_b\!\!\!$&$\!\!\! P_c\!\!$ &\!\!$v_c$\!\!&\!\!$v_b$\!\!& \!\!$\gamma_c$\!\!  \\
$\!\!10^{17}\frac{Kg}{m^3}\!\!\!$&$\!\!10^{17}\frac{Kg}{m^3}\!\!\!$&$\!\!10^{34}Pa$\!\! & \!\!$c$\!\!& \!\! $c$&\!\!   \\  \hline
\!8.126410 & 3.086713 &\!\! 1.278207 \!\!\!& 0.4976045\!& 0.6082809&\!\! 1.662451 \!\!\\  \hline
\!12.83818 & 3.717407 &\!\! 2.420873 \!\!\!& 0.4998542\!& 0.6528634&\!\! 1.440711 \!\!\\  \hline
\end{tabular}
\caption{Values of the hydrostatic variables in the center and on the surface for the minimum and maximum compactness.}
\label{Tabla}
\end{center}
\end{table}
The values of the central and border densities obtained from the theoretical model constructed considering the minimum and maximum compactness values for the neutron star
PSR J0348+0432 are consistent with orders of magnitude, greater than the nuclear density $\rho_n=2.7\times 10^{17}Kg/m^3$, associated with neutron stars.
The table shows that for greater compactness the density is greater than the central pressure, as expected for this type of objects.
\section{Conclusions} \label{VI}
A solution to Einstein's equations has been presented that can be used to describe compact objects with compactness ratio  $u\leq 0.2660858316$. The solution depends on a parameter that has been algebraically and graphically restricted by imposing the conditions that determine whether the model is physically acceptable or not. We have shown through graphical analysis of the behavior of pressure, density, speed of sound and adiabatic index that the solution is physically acceptable.
Also as an application of the model, we describe the neutron star PSR J0348+0432 obtaining that the maximum density value $1.283818 \times 10^{18} Kg/m^3$ it happens for the value
of the mass  $2.05 M_\odot$ and radio $12.065 Km$. The model could also be useful to describe more compact objects to the neutron star PSR J0348+0432 and according to the obtained these
stars would be more dense. This solution leads to future work in which we can consider generalizations of the model for the case of anisotropic or charged stars, as has been approached in relation to other solutions \cite{Tolman,Singh} to describe other compact stars. Or, the possibility of describing the neutron star  PSR J0348+0432 using a charged or anisotropic stellar model and through a comparison between the possible models and based on observational data, discerning about the convenience of some of these models on the model presented with the interior described by a perfect fluid, questions that could be made later.

\section*{ACKNOWLEDGMENTS}
We appreciate the facilities provided by the Michoacana  de San Nicolás de Hidalgo  University  during the conduct of this research.
%and we appreciate the comments and suggestions made by the reviewer that have allowed the presentation of results with greater clarity.

\end{document}